\documentclass[pdflatex,sn-basic]{sn-jnl}
\usepackage{graphicx}
\usepackage{amsmath,amssymb}
\usepackage{booktabs}
\usepackage{xcolor}
\newcommand{\papertitle}{Can open citation data detect citation stacking? An audit against Clarivate's journal suppressions}
\newcommand{\papershorttitle}{Open citation data and citation stacking}
\newcommand{\paperkeywords}{citation stacking, journal self-citation, Journal Impact Factor, OpenCitations, research integrity, anomaly detection}
\begin{document}

\title[\papershorttitle]{\papertitle}

\author*[1]{\fnm{Hossein} \sur{Mohasel Arjomandi}}\email{hosseinm@microsoft.com}
\equalcont{The views expressed in this article are the author's own and do not necessarily reflect those of Microsoft.}
\affil*[1]{\orgname{Microsoft}, \orgaddress{\city{Redmond}, \state{WA}, \country{USA}}}

\abstract{Clarivate suppresses journals from the Journal Citation Reports when self-citations or concentrated citations from
another journal (``citation stacking'') distort their Journal Impact Factor. Its data are closed. Yearly
journal citation networks were built from the OpenCitations snapshot of January 2024 (1.96 billion citations; 85,437 journals)
and compared with Clarivate's suppression lists for data years 2015--2023, collected from primary sources. Where
reference lists are present, open data reproduce Clarivate's published self-citation shares within ten
points for 36 of 50 journals. The failures follow publishers that do not deposit references with Crossref, which also
hides several stacking donors. Biodiversity data downloads registered with DataCite inflate naive impact-factor
numerators of taxonomy journals. The leading open method, CIDRE, detects none of 15 named
Clarivate pairs: it drops large donor journals, and lowering its donor threshold merges the pairs into large groups. A recipient-centric reading of CIDRE's own test flags stacking-listed journals 6 to 14 times more often than
other journals, depending on consensus level, and 5 to 7 times more often than size-matched controls, sometimes years
before the listing. At equal flag rate it finds more listed journals than a plain 43\% donor-share rule (10 vs 3 of
28), and about as many as a rule with a citation floor. A prospective check on the 2024--2025 lists agrees in
direction but is not significant (one of six). Flagged pairs show more citations per citing paper and weaker topical
similarity than matched pairs. Open audits are feasible if coverage gaps are reported.}


\maketitle

\section{Introduction}\label{sec:intro}

The Journal Impact Factor (JIF) counts the citations a journal receives in one year to the items it published in the
two previous years. Because the JIF still shapes hiring, funding and journal choice, some journals inflate it, either
by citing themselves or by receiving concentrated citations from a partner journal, a practice known as citation
stacking \citep{heneberg2016stacking,vannoorden2013brazil,wilhite2012coercive,chorus2016coercive}. Each year Clarivate
withholds the JIF of journals whose citation patterns it judges anomalous and publishes the list, with key figures
such as the share of self-citations in the JIF numerator or the share of a recipient's numerator that comes from one
donor \citep{moussa2022suppressed,taskin2021selfcitation}.

These decisions rest on Web of Science data, which outsiders cannot inspect. Open citation data now make independent
checks possible in principle: OpenCitations releases citation links under open licences
\citep{peroni2020opencitations,heibi2019coci}, and its Meta collection gives the bibliographic records behind them
\citep{massari2024meta}. Whether these data are good enough for integrity work is less clear. Coverage comparisons
show that open sources approach the large commercial indexes on average
\citep{martinmartin2021coverage,visser2021comparison,culbert2025openalex}, but an average says little about the small,
specific sets of journals that suppressions concern.

The best-known open method for this task is CIDRE \citep{kojaku2021cidre}. It removes citations that a degree-corrected
stochastic block model can explain, scores journals as donors and recipients of the remaining ``excessive''
citations, and keeps groups of journals that pass a threshold. On Microsoft Academic Graph data it found 12 of 22
groups that Clarivate had suspended, and community detection found none.

Three questions are addressed. First, how faithfully do fully open data reproduce the figures Clarivate publishes, and where
and why do they fail? Second, does CIDRE, run as published, detect the stacking cases Clarivate listed after its
own validation period, and if not, why? Third, what do the excess citations look like at the level of individual
papers? The contributions are:
\begin{itemize}
  \item a primary-source label set of Clarivate suppressions for data years 2015--2023 and an audit of open-data
        fidelity against Clarivate's published self-citation and stacking figures, with the failures traced to
        publishers' reference deposit and to a data-citation artifact;
  \item a diagnosis of a structural blind spot in CIDRE: its group-building step drops large donor journals, so the
        pattern Clarivate most often reports, a large journal feeding a small one, cannot form a group;
  \item a recipient-centric reading of CIDRE's own excessive-edge test that flags listed stacking journals well above
        chance, also against size-matched controls and plain baselines, sometimes years before the listing, with a
        pre-specified evidence check of its flags and a prospective check on the 2024--2025 lists;
  \item an analysis of the sensitivity of CIDRE to its stochastic null-model partition, and a paper-level anatomy of
        flagged journal pairs.
\end{itemize}
All analyses run on public data; code, run logs and checksums are released (Section~\ref{sec:availability}).
\section{Related work}\label{sec:related}

\paragraph{Journal self-citation and citation stacking.}
Coercive and excessive journal self-citation have been documented by surveys and large-scale analyses
\citep{wilhite2012coercive,chorus2016coercive,taskin2021selfcitation}. Heneberg traced how excessive self-citation can
shift into citation stacking between journals \citep{heneberg2016stacking}, and a Brazilian citation scheme drew wide
attention to stacking in 2013 \citep{vannoorden2013brazil}. Repeated suppressions from the Journal Citation Reports
have been analysed bibliometrically \citep{moussa2022suppressed}.

\paragraph{Detecting anomalous citation groups.}
\citet{fister2016cartels} proposed searching citation networks for cartels, and unsupervised anomaly
detection has been applied to journal-level networks \citep{jolly2020anomaly}. CIDRE \citep{kojaku2021cidre} is the
reference method for journal groups validated against Clarivate's decisions. Paper-level approaches use graph learning
\citep{liu2022tnnls}, and recent work combines journal and paper levels in heterogeneous graphs \citep{liu2026isf}.
\citet{barnes2026distance} show that journals flagged in the Journal Citation Reports receive citations from closer parts of the
journal network.

\paragraph{Open data.}
Coverage of OpenCitations and other open sources has been compared with commercial indexes
\citep{martinmartin2021coverage,visser2021comparison,culbert2025openalex}, but not, to the author's knowledge, in terms of what it
means for detecting citation manipulation. Concurrent work by \citet{zhang2026openalex}, released as a dataset
with code, compares 15 detection methods on OpenAlex for 1,057 journals in chemistry, materials science and
nanoscience, using 16 Clarivate stacking pairs from 2018--2025. The present study differs in data source (OpenCitations), scope (all
fields), the fidelity audit against Clarivate's published figures, the coverage diagnosis, and the analysis of
CIDRE's mechanism.
\section{Data and methods}\label{sec:methods}

\subsection{Citation data and journals}
The data are the parsed OpenCitations snapshot of January 2024 released through the Illinois Data Bank
\citep{arjomandi2024dataset}, with 1,962,840,983 citation links between integer-identified publications. Journal
(venue) assignments come from the OpenCitations Meta CSV dump \citep{ocmeta2024dump,massari2024meta}, joined on the
publication identifier; venue records are merged when they share an ISSN. Of 76.8 million journal articles,
72,720,703 obtain their venue from Meta, 388,394 from the venue field of the parsed file, and 3,695,762 have none. A
venue counts as a journal if its own Meta record has type \emph{journal} and it has an ISSN; this rule keeps 92,779
venues, of which 85,437 appear in the citation network. A stricter rule that also requires at least half of the
venue's items to be journal articles is used as a sensitivity check. Processing used DuckDB \citep{raasveldt2019duckdb}.

\subsection{JIF-window counts and journal scores}
For a citing year $t$, let $c(A\!\to\!B,t)$ be the number of citations from items of journal $A$ published in $t$ to
items of journal $B$ published in $t-1$ or $t-2$, the JIF window. Citing items of any type are counted but, following
Clarivate's practice of counting citations from indexed sources, only those that have a journal; including
citing items without a journal is a sensitivity analysis. For each recipient journal $B$ and year $t$ the following are computed: the numerator $n_B(t)=\sum_A c(A\!\to\!B,t)$, the self-citation share $c(B\!\to\!B,t)/n_B(t)$, a JIF proxy dividing $n_B(t)$
by $B$'s journal articles from $t-1$ and $t-2$, and the share of $n_B(t)$ from its largest external donor.

\subsection{Labels: Clarivate suppressions 2015--2023}
The suppression lists were rebuilt from primary Clarivate material: the ``Suppressed Titles 2007--2017'' spreadsheet
(data years 2015--2017: journal, year and type), the Journal Citation Reports help pages for the 2019, 2022, 2023 and
2024 releases (archived copies), the 2020 release page and a copy of the 2021 release list. The resulting table has 144
rows. There are 98 self-citation rows, 60 of them with a published self-citation share. There are 37 stacking rows:
17 named recipient--donor pairs with Clarivate's two stacking figures, 9 rows for the donors of those pairs, and 11
stacking journals without named partners (2015--2017). Three rows describe one expression of concern (a recipient
with three donors), and six ``suppression warnings'' are excluded. Journals were matched to OpenCitations venues by ISSN taken
from Crossref journal records, with a documented alias or ISSN choice where titles were ambiguous; 131 of 138 names
matched. Table~\ref{tab:accounting} traces each analysis's denominator back to these rows.

The table was checked against Clarivate's current consolidated spreadsheet of suppressed titles (data years
2007--2024), retrieved in October 2026 from the Journal Citation Reports help centre. Every journal listed in that
spreadsheet for 2015--2023 is in the table. Nine self-citation rows taken from the lists published at the time (eight
for data year 2019, one for 2018) are absent from it; they are kept, and results without them are reported as a sensitivity check. For an out-of-sample test, the suppressions for data years 2024 (from the same spreadsheet) and 2025
(from the help-centre page of the 2026 release) were retrieved only after the predictions had been frozen (Section~\ref{sec:m-robust}).

\begin{table}[t]
\centering
\caption{From label rows to analysis denominators.}
\label{tab:accounting}
\small
\begin{tabular}{@{}lr@{}}
\toprule
Step & $n$ \\
\midrule
\multicolumn{2}{@{}l}{\emph{Self-citation fidelity (Section~\ref{sec:res-fidelity})}} \\
\quad self-citation rows & 98 \\
\quad with a published self-citation share & 60 \\
\quad matched to a journal & 59 \\
\quad evaluated ($\ge$20 numerator citations and citable items): references present / absent & 50 / 7 \\
\multicolumn{2}{@{}l}{\emph{Self-citation ranks}} \\
\quad matched self-citation rows & 95 \\
\quad ranked ($\ge$20 numerator citations and citable items) & 85 \\
\multicolumn{2}{@{}l}{\emph{Stacking fidelity}} \\
\quad named pairs with a published donor share (17 pairs and the expression of concern) & 18 \\
\quad both journals matched: donor references present / absent & 9 / 4 \\
\multicolumn{2}{@{}l}{\emph{CIDRE detection (Section~\ref{sec:res-cidre})}} \\
\quad named recipient--donor rows 2015--2023, incl.\ expression of concern & 20 \\
\quad both journals matched & 15 \\
\multicolumn{2}{@{}l}{\emph{Enrichment (Section~\ref{sec:res-variant})}} \\
\quad stacking-listed journals in the base, design cases excluded & 28 \\
\quad self-citation-listed journals in the base & 85 \\
\multicolumn{2}{@{}l}{\emph{Out-of-sample test (Section~\ref{sec:res-oos})}} \\
\quad journals newly listed for stacking, data years 2024--2025 & 8 \\
\quad in the evaluated population & 6 \\
\bottomrule
\end{tabular}
\end{table}

\subsection{Fidelity measures}
Clarivate reports, for self-citation cases, the percentage of self-citations in the JIF numerator and, for stacking
pairs, the donor's percentage of the recipient's JIF numerator and the percentage exchange to JIF years, defined as
``the proportion of all citations from donor to recipient (all years) that reference the two years considered in the
Journal Impact Factor calculation''. The same quantities are computed here; for the exchange share the denominator counts
cited years before $t$, so the estimate is an upper bound. It was fixed in advance that a figure counts as reproduced if it lies within $\pm10$ percentage points,
and that a listed journal ``ranks'' if it is in the top 1\% of its year on the relevant share.

\subsection{CIDRE}
CIDRE was implemented as described by \citet{kojaku2021cidre}, by wrapping the authors' package and closing
three gaps between package and paper: condition 1 (more than half of the $A\!\to\!B$ citations in year $t$ go to the
JIF window), the minimum of $\theta_w=50$ within-group citations, and a compatibility fix. An edge is excessive if it
meets condition 1 and its weight is significant under a Poisson test against the expected weight from a
degree-corrected stochastic block model, with Benjamini--Hochberg correction at $\alpha=0.01$
\citep{benjamini1995}. A journal's donor (recipient) score is the share of its outgoing (incoming) citations that are
excessive within the current group; journals whose donor and recipient scores are both below $\theta=0.15$ are
removed iteratively, and the remaining weakly connected components are the groups. The block model was fitted once on
the network aggregated over all years with graph-tool's nonparametric inference
\citep{peixoto2017nonparametric,peixoto2014graphtool}. Detection follows the paper's rule: a Clarivate group is detected
if at least half of a CIDRE group's journals and at least two journals are shared.

The block-model fit is stochastic. Seven partitions were therefore fitted of the main journal network (five seeds, a
repeat of the configuration seed, and a single-threaded fit), plus one of the stricter journal rule, and CIDRE was run on every year from 2015 to 2023 with each.

\subsection{A recipient-centric reading of CIDRE}
Clarivate's stacking criterion is recipient-centric: a recipient that depends on one donor for a large share of its
numerator. CIDRE's excessive-edge test can be read the same way, without the group step. For journal $j$ in year
$t$, let $e_j$ be the number of excessive citations it receives (self-citations excluded) and $s_j$ its in-strength.
Journal $j$ is flagged if $e_j/s_j \ge \theta$ and $e_j \ge \theta_w$, with CIDRE's own $\theta=0.15$ and $\theta_w=50$. This
variant was designed after CIDRE's misses had been diagnosed on two 2023 cases (Section~\ref{sec:res-cidre}); those two
journals are therefore excluded from every evaluation in which they could favour it. Because flags vary between
partitions, consensus flags present in at least $k$ of the seven partitions are also reported. The variant is evaluated by enrichment: the share of listed journals ever flagged in 2015--2023 compared with the share of all journals with at
least 20 numerator citations in some year, with a one-sided binomial test.

\subsection{Baselines, size matching and an out-of-sample test}\label{sec:m-robust}
These checks were added after a review of a first draft; their protocols were recorded in the lab notebook
before any of their results were computed.

\emph{Baselines.} A plain rule without a null model flags journal $j$ in year $t$ if its largest external donor
provides at least a share $X$ of its numerator (and the numerator is at least 20). $X$ is set on all journals, without
labels, so that the rule flags the same share of journals as the variant; as a second version, CIDRE's two thresholds are applied to the plain share (share $\ge0.15$ and at least 50 citations from the top donor). To isolate the role of
the block model, the variant's test is also run against a single block, that is, a configuration-model null. Rules are
compared on the same listed journals with exact McNemar tests.

\emph{CIDRE with an asymmetric threshold.} CIDRE is rerun with separate thresholds for the two scores: the recipient
threshold kept at $\theta_r=0.15$ and the donor threshold lowered to $\theta_d\in\{0.10,0.05,0.02,0.01\}$, so that a
journal is removed only if its donor score is below $\theta_d$ and its recipient score is below $\theta_r$.

\emph{Size.} Flag rates depend on journal size, so listed journals are also compared with journals of similar size,
measured by the largest yearly numerator, using a Mantel--Haenszel analysis over size deciles and ten never-listed
controls per listed journal of nearest size, drawn from the same block where possible. Confidence intervals for the
matched enrichment come from a cluster bootstrap over listed journals (10,000 resamples).

\emph{Out of sample.} Before Clarivate's lists for data years 2024 and 2025 were opened, the set of journals flagged in
2015--2023 was frozen (a file with a recorded checksum). Among journals with at least 20 numerator citations in some
year and not listed in 2015--2023, flagged journals are then compared with the others on how often they were newly
listed for stacking (one-sided Fisher exact test). The citation data end with data year 2023, so this tests whether a flag
anticipates a later listing.

\subsection{Paper-level anatomy and evidence check}
For flagged recipients and their largest excessive donor, and for named Clarivate pairs, the citing papers behind the pair's
window citations were collected, and the citations per citing paper, the Gini coefficient of citations
over citing papers, and the shares coming from the single top citing paper and from the top 10\% of citing papers were computed. Titles of citing and cited papers were compared by TF--IDF cosine similarity. Each pair receives up to three controls from
the same year with 0.8 to 1.25 times its window citations, both journals in the journal rule, and a recipient never
flagged. As a structured evidence check, fixed before the sample was drawn, three public indicators are recorded per
pair: either journal appears in Clarivate's lists; recipient and donor share a Crossref publisher; or the top 10\% of
citing papers carry at least half of the pair's citations.

\subsection{Reproducibility}
Every analysis is a script that writes a run folder with its parameters, input and output checksums and log, and the
decisions and pass criteria were recorded in a lab notebook before the corresponding runs. Pre-specified and post hoc analyses are distinguished throughout.
\subsection{Use of generative AI}\label{sec:ai-use}
Claude Opus 5.5 (Anthropic) was used as a coding and writing assistant: to write and run analysis scripts under the
author's direction, to search the literature, to draft and edit the text, and to review drafts. All analyses, numbers
and references were checked by the author against the run records and the original sources (DOIs), and the author
takes full responsibility for the content.
\section{Results}\label{sec:results}

\subsection{What open data contain}\label{sec:res-coverage}
\begin{figure}[t]
\centering
\includegraphics[width=0.6\linewidth]{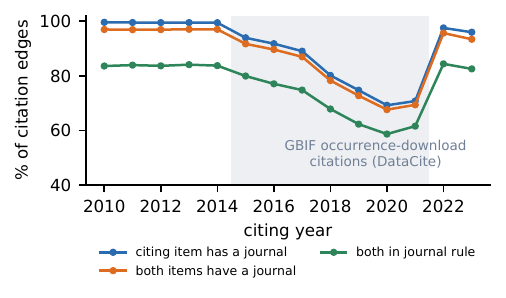}
\caption{Share of citation links whose citing item has a journal, whose both items have a journal, and whose both
items are journals under the journal rule, by citing year. The 2015--2021 dip is due to GBIF occurrence-download
records registered with DataCite.}
\label{fig:coverage}
\end{figure}

In citing years up to 2014, 99.5\% of citation links have a citing item with a journal; this share falls to 69.3\%
in 2020 and recovers to 97.6\% in 2022 (Fig.~\ref{fig:coverage}). The dip is caused by one source. In 2020, 221,918 of
the 230,438 citing items typed \emph{dataset} without a journal (96.3\%) carry DOIs of the Global Biodiversity
Information Facility (prefix 10.15468, registered with DataCite): occurrence downloads whose reference lists cite the
datasets and taxonomic publications the records come from. These are legitimate data citations, but they are not the
journal citations Clarivate counts. Had they been counted, they would form 12.8\%, 14.4\% and 11.2\% of the JIF-window
numerator in 2019, 2020 and 2021, concentrated in a few taxonomy journals: in 2020, 99.8\% of the numerator of
\emph{Zootaxa} would come from such items, giving a JIF proxy above 600. Restricting citing items to those with a
journal removes the artifact; including them changes the JIF proxy noticeably only in 2019--2021 (correlation with the
main version 0.70--0.90, against at least 0.99 in other years).

\subsection{Fidelity to Clarivate's published figures}\label{sec:res-fidelity}
\begin{figure}[t]
\centering
\includegraphics[width=0.55\linewidth]{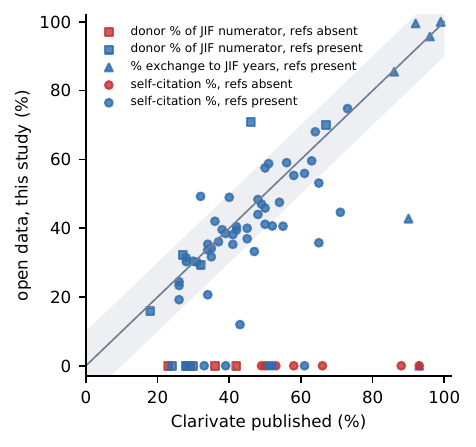}
\caption{Clarivate's published percentages against ours, by whether the citing journal's (self-citation) or donor's
(stacking) reference lists are present in OpenCitations. The band marks $\pm10$ percentage points.}
\label{fig:fidelity}
\end{figure}

Whether open data reproduce Clarivate's figures depends mainly on whether the citing journal's reference lists are in
OpenCitations (Fig.~\ref{fig:fidelity}). When they are, the open-data self-citation share lies within $\pm10$ percentage points
of Clarivate's for 36 of 50 journals, with a median absolute difference of 4.6 points and a median signed difference
of $-2.8$ points. When they are absent, the open-data share is zero and none of seven journals passes. For stacking pairs whose
donor has references and the pair has citations in the open data (five pairs), the donor's share of the recipient's
numerator is reproduced for four and the exchange share for four. Many stacking donors in Clarivate's lists, however,
have no references in OpenCitations at all.

The missing references are not random. For the journals whose references are absent, the median share of their DOIs
with deposited reference lists in Crossref is zero (zero for eight of eleven), against 0.96 for journals whose
references are present. The journals of one publisher that deposits references today began only recently (backfile
coverage around 5\%). These blind spots include donors in several Clarivate stacking cases.

On ranks, 59 of 85 self-citation suppressions (69\%) lie in the top 5\% of their year by self-citation share, with a
median percentile of 3.1\%, but only 9 are in the top 1\%. The criterion fixed in advance for data year 2020,
at least half of the suppressed journals in the top 1\%, was not met (0 of 8) and is reported as recorded.
Without the nine self-citation rows that are missing from Clarivate's current spreadsheet, 55 of 76 journal-years are
in the top 5\% (72\%) and 9 in the top 1\%, and 28 of 41 journals with references present are reproduced within
$\pm10$ points.

\subsection{CIDRE on open data}\label{sec:res-cidre}
CIDRE finds between 11 and 32 groups per year, typically of three to five journals. It detects none of the 15 named
Clarivate pairs from 2018--2023 for which both journals are matched, in any of the eight partitions, and no recipient
or donor of these pairs is in any CIDRE group in its year. Some misses are due to data: the donor has no references in
OpenCitations, or the pair has too few citations to reach $\theta_w$. Two 2023 pairs, however, are well covered and
still missed.

\begin{figure}[t]
\centering
\includegraphics[width=0.6\linewidth]{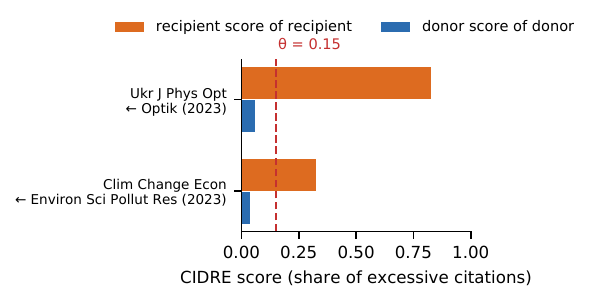}
\caption{Why CIDRE misses two well-covered 2023 stacking pairs: the recipients' scores exceed $\theta$, but the large
donors' scores do not, so the group-building step removes the donors.}
\label{fig:peeling}
\end{figure}

For these, a stage-by-stage diagnosis shows the mechanism (Fig.~\ref{fig:peeling}). The stacking edge passes CIDRE's
test: 99.5\% (Optik to Ukrainian Journal of Physical Optics) and 85.5\% (Environmental Science and Pollution Research
to Climate Change Economics) of the donor's citations to the recipient fall in the JIF window, and the edges are
significant against the block model. The recipients' scores, 0.83 and 0.32, are well above $\theta$. But the donors are
large journals whose stacking citations are a small part of their output: their donor scores are 0.057 and 0.035.
Since CIDRE removes a journal when both of its scores are below $\theta$, the donor is removed, the recipient's score
falls to zero, and no group remains. The CIDRE paper motivates $\theta=0.15$ on the recipient side, noting that all
recipients suspended by Clarivate received at least 15\% of their citations from donors \citep{kojaku2021cidre}, but
applies the same threshold to donors.

Lowering the donor threshold alone does not repair this. With $\theta_r=0.15$ and
$\theta_d=0.10$, 0.05, 0.02 or 0.01, CIDRE still detects none of the 15 pairs. Keeping large donors joins their components:
the largest group has 28 journals at the published setting, and 29, 163, 656 and 3,056 journals at
$\theta_d=0.10$, 0.05, 0.02 and 0.01. The named pairs then sit inside groups much larger than the pair, and the overlap rule fails. At $\theta_d=0.05$,
Ukrainian Journal of Physical Optics and Optik share a group of 13 journals; at $\theta_d=0.02$, three 2023 pairs share one group
of 345 journals. The blind spot thus lies in how groups are formed, not only in the value of the threshold.

CIDRE does find stacking between small journals. In 2015 one of its groups consists exactly of Applied Clinical
Informatics and Methods of Information in Medicine, and in 2017 one consists exactly of Liver Cancer and Digestive
Diseases; in each case both journals were suppressed for stacking in that year, although Clarivate's spreadsheet for
those years does not name the pairs.

\subsection{The recipient-centric reading}\label{sec:res-variant}
Over 2015--2023, the variant flags 12 of 30 stacking-listed journals (40\%), against 3.4\% of all journals with at
least 20 numerator citations in some year: an enrichment of 11.7 ($p=1.2\times10^{-10}$). Without the two journals
used to design it, 10 of 28 are flagged (10.5-fold, $p=1.6\times10^{-8}$), and across partitions and size thresholds
the enrichment ranges from 5.7 to 10.5 ($p\le1.1\times10^{-5}$). For self-citation-listed journals the enrichment is
weaker and not robust across partitions (1.3 to 3.4), as expected for a measure of external citations.

\begin{figure}[t]
\centering
\includegraphics[width=0.6\linewidth]{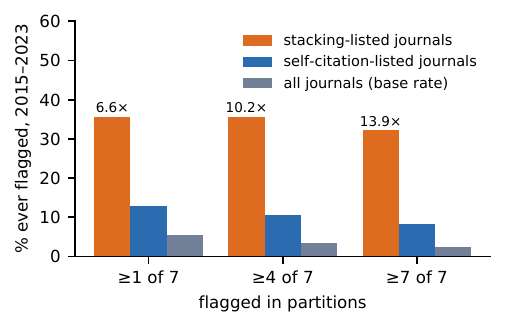}
\caption{Share of listed journals ever flagged by the recipient-centric reading in 2015--2023, against the base rate,
by the number of partitions required (design cases excluded). Labels give the enrichment for stacking-listed journals.}
\label{fig:enrichment}
\end{figure}

Requiring flags to appear in more partitions lowers the base rate faster than the hit rate (Fig.~\ref{fig:enrichment}).
With flags present in all seven partitions, 2.3\% of journals are ever flagged, against 9 of 28 stacking-listed
journals: a 13.9-fold enrichment ($p=8.7\times10^{-9}$). Because the consensus rule was chosen after seeing these
results, all three levels are reported.

Two checks ask whether simpler explanations account for this enrichment (Table~\ref{tab:baselines}). The first is a
plain rule without a null model. Set to flag as many journals as the variant, it requires the largest donor to provide
at least 43\% of the numerator, and it flags 3 of 28 stacking-listed journals (3.1-fold, 95\% CI 1.1--8.0, $p=0.069$).
The variant reaches 10.5-fold (6.1--15.9) at the same flag rate. The variant's test against a single block instead of
the block model flags 3.4 times as many journals, and its enrichment falls to 4.3-fold (2.8--5.8). The block model thus
makes the test more specific. On the same 28 journals, the variant flags 8 that the rate-matched rule misses, and the
rule flags 1 that the variant misses (exact McNemar test, $p=0.039$). A plain rule with CIDRE's two thresholds applied to
the raw share comes closer, at 8.4-fold (4.3--14.5). It flags 7 listed journals; the variant flags 4 that it misses and
it flags 1 that the variant misses ($p=0.38$). An absolute volume floor therefore carries much of the signal. Overall
the two rules' flags overlap little (Jaccard 0.22), so they find partly different journals.

The second check concerns size. Stacking-listed journals are not small: 25 of the 28 lie in the upper six deciles of
journal size (largest yearly numerator). The variant needs at least 50 excessive citations, so it flags no journal in
the three lowest deciles and 7--9\% of journals in the highest. Enrichment against all journals therefore partly
reflects size. Against ten never-listed controls of nearest size per listed journal, 71\% of them from the same block,
the variant flags 35.7\% of listed journals and 7.1\% of controls: 5.0-fold (95\% CI 2.7--10.9, cluster bootstrap over
listed journals; $p=6.8\times10^{-5}$, Fisher's exact test). The consensus flags 32.1\% against 4.6\%: 6.9-fold
(3.1--17.1, $p=2.6\times10^{-5}$). Over size deciles the Mantel--Haenszel odds ratios are 11.3 and 15.0. Size matching
moves the rate-matched rule the other way, from 3.1- to 5.0-fold. A 43\% share is easy to reach with a small
numerator, so this rule flags 6\% of journals in the lowest size decile but 0.2\% in the highest. Size-matched
controls of the larger listed journals are therefore rarely flagged. With only 3 listed journals and 6 controls flagged,
its interval is wide (0--30). For self-citation-listed journals the variant's size-matched enrichment is 1.7-fold and not
significant ($p=0.09$).

\begin{table}[t]
\centering
\caption{Enrichment of stacking-listed journals (28, design cases excluded) ever flagged in 2015--2023, by flagging
rule. Flag rate: share of the 29,611 journals with at least 20 numerator citations in some year that are ever flagged.
Size-matched: against ten never-listed controls of nearest size per listed journal (95\% CI from a cluster bootstrap
over listed journals).}
\label{tab:baselines}
\footnotesize
\setlength{\tabcolsep}{4pt}
\begin{tabular}{@{}lrrrr@{}}
\toprule
Rule & Flag rate & Listed flagged & Enrichment (95\% CI) & Size-matched (95\% CI) \\
\midrule
Variant, partition \emph{main} & 3.4\% & 10 & 10.5 (6.1--15.9) & 5.0 (2.7--10.9) \\
Variant, consensus of 7 partitions & 2.3\% & 9 & 13.9 (7.8--21.9) & 6.9 (3.1--17.1) \\
Variant, single-block null & 11.6\% & 14 & 4.3 (2.8--5.8) & -- \\
Top-donor share $\ge0.43$ (rate-matched) & 3.4\% & 3 & 3.1 (1.1--8.0) & 5.0 (0--30.0) \\
Top-donor share $\ge0.15$ and $\ge50$ citations & 3.0\% & 7 & 8.4 (4.3--14.5) & 5.0 (1.7--12.9) \\
\bottomrule
\end{tabular}
\end{table}

\begin{figure}[t]
\centering
\includegraphics[width=0.6\linewidth]{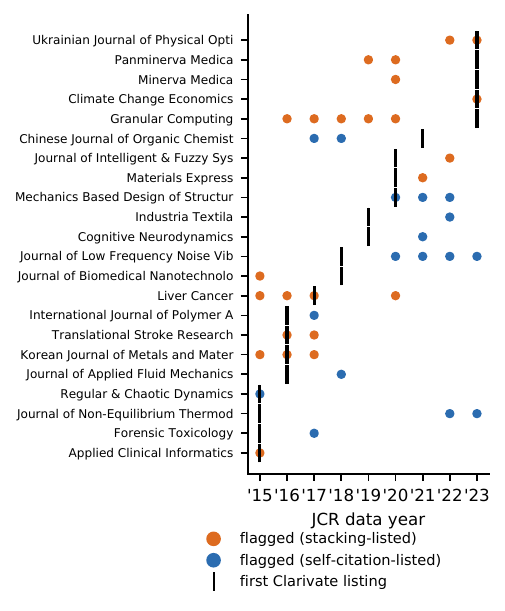}
\caption{Years in which listed journals were flagged (dots) and their first Clarivate listing (bars), partition
\emph{main}.}
\label{fig:leadtime}
\end{figure}

Eight listed journals were flagged before their first listing (Fig.~\ref{fig:leadtime}). Granular Computing was
flagged in every year from 2016 to 2020, three to seven years before its 2023 listing, with Information Sciences, its
donor in Clarivate's list, as the largest excessive donor in four of those years. Panminerva Medica and Minerva Medica
were flagged three to four years before their 2023 listing, Liver Cancer one to two years before 2017 (largest
excessive donor Digestive Diseases), and Ukrainian Journal of Physical Optics in 2022 (donor Optik).

In the pre-specified evidence check on 196 randomly sampled consensus flags and 196 matched controls, 28.1\% of flagged
pairs carry at least one independent indicator against 6.1\% of controls ($p=6.2\times10^{-9}$, Fisher's exact
test). The most frequent indicator is a shared publisher of recipient and donor (19.4\% against 6.1\%); 5.1\% of
flagged pairs have a donor in Clarivate's lists and 5.6\% have strongly concentrated citations, against none of the
controls. These shares are a lower bound on the precision of the flags, not an estimate of it.

\subsection{A prospective check on the 2024 and 2025 lists}\label{sec:res-oos}
Clarivate listed eight journals for stacking in data years 2024 and 2025, six of which are in the evaluated
population. One of them, Asian Journal of Agriculture and Biology, was among the frozen consensus flags (flagged in
2022) and was listed for data year 2024. Thus 1 of 6 newly listed journals was flagged, against 2.3\% of the other
29,490 journals (7.4-fold, one-sided Fisher $p=0.13$). With the flags of partition \emph{main} the result is 5.0-fold
($p=0.19$). The direction agrees with the retrospective results, but with six cases the test has little power and the
result is not significant. A flag can anticipate a later listing only if the pattern persists beyond the citation data, which
end with data year 2023.

\subsection{Paper-level anatomy}\label{sec:res-anatomy}
\begin{figure}[t]
\centering
\includegraphics[width=0.6\linewidth]{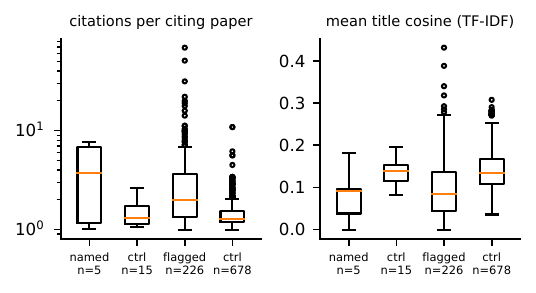}
\caption{Citations per citing paper (log scale) and citing--cited title similarity for named Clarivate pairs, flagged
pairs and their matched controls.}
\label{fig:anatomy}
\end{figure}

Flagged pairs differ from matched controls in how their citations arise (Fig.~\ref{fig:anatomy}). Each citing paper
cites the recipient 1.98 times on median, against 1.28 ($p=2.5\times10^{-27}$, Mann--Whitney), and citations are more
unevenly spread over citing papers (Gini 0.25 against 0.18, $p=1.5\times10^{-8}$). The share from the top 10\% of
citing papers does not differ ($p=0.35$): the pattern is many papers each citing a little too often rather than a few
papers with long reference lists. Cited and citing titles are less similar in flagged pairs (TF--IDF cosine 0.084
against 0.133, $p=1.2\times10^{-24}$). Since excessive edges are defined against the block model, flagged pairs might
simply join more distant fields; in fact 27.9\% of flagged pairs share a block against 0.6\% of controls, and within
cross-block pairs the differences remain (cosine 0.084 against 0.133, $p=1.2\times10^{-17}$; citations per citing
paper 2.00 against 1.28). For the five named Clarivate pairs with citations the sample is too small for conclusions.

\subsection{Sensitivity to the block-model partition}\label{sec:res-stability}
\begin{figure}[t]
\centering
\includegraphics[width=0.6\linewidth]{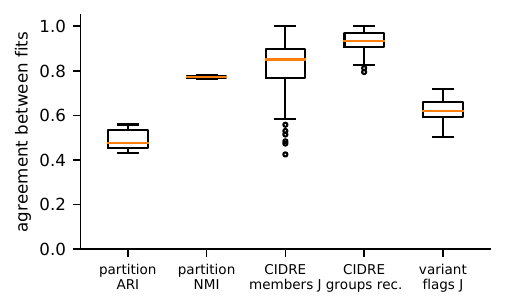}
\caption{Agreement between fits: block-model partitions (ARI, NMI), CIDRE group members (Jaccard) and group recovery,
and the variant's flags (Jaccard).}
\label{fig:stability}
\end{figure}

With eight threads, block-model fits are not reproducible from a seed: six fits of the same network give 2,131 to 2,912
blocks, with pairwise normalized mutual information of 0.76--0.78 and adjusted Rand index \citep{hubert1985ari} of
0.44--0.54. Two single-threaded fits with the same seed give identical partitions, so the variation comes from the
parallel search. CIDRE's groups depend much less on the partition than the partitions themselves agree: the median
Jaccard similarity of group members between partitions is 0.85, and 93\% of groups have a counterpart (Fig.~\ref{fig:stability}).
The variant's flags are less stable (median Jaccard 0.62), which motivates the consensus rule.

\section{Discussion}\label{sec:discussion}

\paragraph{What open data can and cannot show.}
Open citation data reproduce Clarivate's self-citation figures closely when reference lists are present, so open
audits of journal citation behaviour are feasible. But the gaps sit where integrity questions arise: some publishers
deposit no references, and several of the donors behind Clarivate's stacking cases are among them. An open audit can
therefore not conclude that a journal shows no anomaly when its citing partners are absent from the data; coverage
of the relevant journals should be reported alongside any result. Data citations from repositories, such as GBIF
downloads, are a second, newer source of distortion for anyone computing impact-factor-like indicators from open
data.

\paragraph{Why CIDRE misses Clarivate-style stacking.}
CIDRE was designed to find groups whose members exchange excessive citations, and it does find stacking between small
journals. Its symmetric threshold, however, requires a donor to devote a substantial share of its own citations to the
group. In the stacking cases Clarivate reported in 2023, large journals fed small ones, and CIDRE flags the edge but
discards the group. This is a property of the method, not of the data: it appears with well-covered donors, and
lowering the donor threshold merges the pairs into large groups instead of isolating them. A recipient-centric reading
of the same test recovers listed stacking journals far above chance and sometimes before listing, at the cost of
flagging about 2--5\% of journals over nine years, depending on the consensus level. At the same flag rate it
finds more listed journals than a plain rule on the top donor's share (10 against 3 of 28). Part of the raw enrichment
reflects journal size, but five- to seven-fold enrichment remains against size-matched controls. A plain rule with a
volume floor finds 7, not significantly fewer, and flags different journals, so the two are better seen as
complementary screens than as rivals.

\paragraph{What the flags mean.}
The flags mark anomalous citation patterns, not misconduct. Many have benign explanations, such as a society journal
and its companion, editorial preferences, or regional communities. The evidence check shows that flagged pairs carry
independent indicators far more often than matched pairs, and the paper-level anatomy shows that their citations are
spread over many papers that cite the recipient repeatedly and are topically further from it. Flags are candidates for
expert review, not verdicts.

\paragraph{Limitations.}
Clarivate names few stacking pairs, so recall figures rest on a small number of cases, and the enrichment results rely
on 28 to 30 stacking-listed journals. The recipient-centric reading and its consensus rule were chosen after part of
the evaluation. The cases used in the design were excluded and all variants are reported. The prospective check on the 2024
and 2025 lists points the same way but rests on six journals and is not significant; a larger test will become possible
as Clarivate publishes further lists and open data extend beyond 2023. The label table includes nine self-citation rows
from the lists published at the time that Clarivate's current spreadsheet omits; results without them are similar.
The asymmetric-threshold test of CIDRE was evaluated on the 15 named pairs and on journal-level enrichment only.
The evidence check is a lower bound on precision. Clarivate's exact formulas and thresholds are not
public, and the exchange share computed here is an upper bound. Crossref coverage statistics describe the present, not the state at
the time of the snapshot. Title similarity uses TF--IDF on titles only, and the anatomy controls are matched on year and volume, not on
field; the size-matched enrichment uses block membership only as a proxy for field.
\section{Conclusions}\label{sec:conclusions}
Fully open citation data can reproduce much of what Clarivate publishes about its journal suppressions, and they can
anticipate some of its stacking decisions. Two conditions matter: auditors must check that the relevant reference
lists are present, and methods must match the pattern they target. A large journal feeding a small one is invisible to
CIDRE's group definition but visible to a recipient-centric reading of the same statistical test. The
label table, code and run records are released so that the audit can be repeated on new suppression lists.

\backmatter

\section*{Declarations}\label{sec:availability}

\begin{itemize}
\item \textbf{Funding:} No funding was received for this work.
\item \textbf{Competing interests:} The author has no competing interests to declare.
\item \textbf{Ethics approval and consent to participate:} Not applicable.
\item \textbf{Consent for publication:} Not applicable.
\item \textbf{Data availability:} Citation data: \citet{arjomandi2024dataset}; venue metadata: \citet{ocmeta2024dump}. The suppression label table
derived from Clarivate's public lists (with source URLs, web-archive links and checksums of the files used; Clarivate's
files themselves are not redistributed), all scripts, configuration, run logs and checksums:
\url{https://github.com/HosseinMA96/citation-stacking-open-data}. The repository is public and will be archived with a DOI on
publication.
\item \textbf{Code availability:} See Data availability.
\item \textbf{Author contributions:} Hossein Mohasel Arjomandi is the sole author and designed the study, ran the analyses and wrote the manuscript.
\end{itemize}

\bibliography{refs}

\end{document}